\documentclass[aps,prl,preprint,groupedaddress,showpacs]{revtex4}
\usepackage{graphicx}

\begin{document}
\title{Structural Vulnerability of the North American Power Grid}

\author{R\'eka Albert$^{1,2}$ Istv\'an Albert$^2$ and Gary L. Nakarado$^3$}
\affiliation{1. Department of Physics, Pennsylvania State University, University Park, PA 16802}
\affiliation{2. Huck Institute for Life Sciences, Pennsylvania State University, University Park, 
PA 16802}
\affiliation{3. National Renewable Energy Laboratory, Golden, CO 80401}

\begin{abstract}

The magnitude of the August 2003 blackout affecting the United States has put 
the challenges of energy transmission and distribution into limelight. 
Despite all the interest and concerted effort, the complexity and interconnectivity 
of the electric infrastructure have so far precluded us from understanding why certain events 
happened. In this paper we study the power grid from a network perspective and determine 
its ability to transfer power between generators and consumers when certain nodes are disrupted. 
We find that the power grid is robust to most perturbations, yet disturbances affecting key 
transmision substations greatly reduce
its ability to function. We emphasize that the global properties of the 
underlying network must be understood as they greatly affect local 
behavior. 
\end{abstract}


\maketitle

During the past decades the North American power infrastructure has evolved into what many experts 
consider the largest and most complex system of the technological age. Geographically, the power 
grid forms a network of over 1 million kilometers of high voltage lines that are continuously 
regulated by sophisticated flow control equipment\cite{roadmap}. As a result of the recent
deregulation of power generation and transmission, about one-half of all domestic 
generation is now sold over ever-increasing distances on the wholesale market before it is 
delivered to customers\cite{roadmap}. Consequently the power grid is witnessing power flows in 
unprecedented magnitudes and directions\cite{rely}. 

As the power grid increases in size and complexity, it is becoming more important to understand 
the emergent behaviors that can take place in the system. Performing an analytic description of the 
electromagnetic processes integrated over the whole grid is a daunting, if not impossible, task. 
Instead the power industry must resort to constructing models that can be used to simulate the network's 
response to various external parameters. Generally these models attempt to simulate actual 
electrical flow characteristics in smaller systems like a single distribution grid\cite{flow}. 
In the present analysis we propose an alternative approach based on recent advances in understanding 
the structure of large complex networks\cite{ab02}. We choose to investigate the network 
representation of the power grid from a topological perspective with the hope of finding 
properties and behaviors that transcend the abstraction.

We have built the network model based on data stored in the POWERmap mapping 
system developed by Platts\cite{platts}, the energy information and market services unit of 
the McGraw-Hill Companies. This mapping system contains information about every power plant, 
major substation and  $115-765$ kV power line of the North American power grid. Our model 
represents the power grid as a network of 14,099 nodes (substations) and 19,657 edges 
(transmission lines). We distinguish three types of substations: generators are the sources
for power, transmission substations transfer the power among high voltage transmission lines, and 
distribution substations are at the “outer edge” of the transmission grid, and the centers of local 
distribution grids. Only the identity of generating substations was directly available from our data 
sources. We identify distribution substations by the criterion of having a single high-voltage 
transmission line connected to them, with the expectation that the flow out of them is continued on
smaller voltage feeder lines leading to consumers\cite{miss}. A total of 1633 nodes are power plants,
we classify 2179 nodes as distributing substations, with the rest being labeled as transmission 
substations.
 
We consider the power from a generator to be accessible to a consumer if there is a path of 
transmission lines between the two. In practice, the existence of 
a connection between two substations does not always imply that power can be transferred across it 
as there may be capacity or other constrains present. By ignoring these our model provides an 
idealized view, a “best case scenario” regarding the characteristics of the grid. We find that the 
network representation of the power grid contains a single connected component, meaning that 
there is a path of transmission lines between any power plant and any distribution substation. 
This observation implies that in the best case scenario each distribution substation can possibly 
receive power from any generator. 

Recent advances in mapping the topology of complex networks have uncovered that a large fraction 
of them are highly heterogeneous with respect to the number of edges incident on a node (also 
called the node degree). In these networks the majority of the nodes have low degrees, but there
is a continuous hierarchy of high-degree nodes (hubs) that play an important role in the system.
The degree distribution of these networks follows a power-law $P(k)\sim k^{-\gamma}$ with the exponent
$\gamma$ mostly between $2$ and $3$. It was demonstrated both numerically and analytically that these
so-called scale-free networks are resilient to the random loss of nodes, but are vulnerable to attacks 
targeting the high-degree hubs\cite{ajb00,cnsw00,cebh01}. Therefore it is important both from a 
theoretical and practical standpoint to determine whether the connectivity of the power grid is reliant
on a small set of hubs and whether their loss will cause a large-scale breakdown of the power grid's 
transmission capability. 

\begin{figure}
\includegraphics[width=8cm,angle=-90]{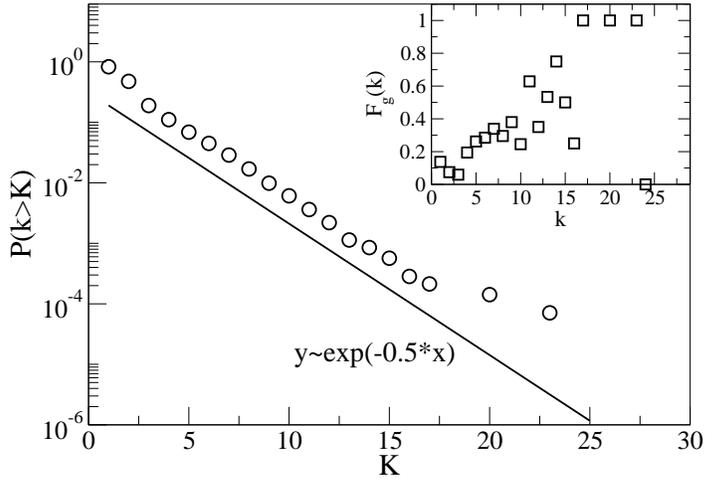}%
\caption{\label{degree_fig} The probability that a substation has 
more than $K$ transmission lines. The straight line represents the exponential 
function (\ref{deg_eq}). Inset: the fraction $F_g(k)$ of generating substations among substations with 
degree $k$.}
\end{figure}

As the node degree is a good indicator of its topological importance, we first determine the degree 
distribution of the power grid. We find that the cumulative degree distribution defined as 
$P(k>K)=\sum_{k>K} P(k)$ follows an exponential  
\begin{equation}
P(k>K)\sim exp(-0.5 K) 
\label{deg_eq}
\end{equation}
(see Fig. \ref{degree_fig}).
This functional form agrees with previous results on the degree distribution of the Western power 
grid\cite{asbs00} and its classification as a single-scale network. The cumulative degree
distribution shows that the probability of high-degree nodes is less than in a scale-free network, 
but higher than in a random network with the same number of nodes and edges. Power engineering
principles suggest that the hubs of the power grid should belong to central station generators, and 
transmission substations should not have more than a few edges. Indeed, the inset to Fig. \ref{degree_fig}
shows that the fraction of generating substations among substations of
a given degree increases with this degree. Surprisingly, however, there are several high-degree 
transmission substations (e.g. $50$ have degree higher than $10$), including the node with highest 
degree. 

\begin{figure}
\includegraphics[width=8cm,angle=-90]{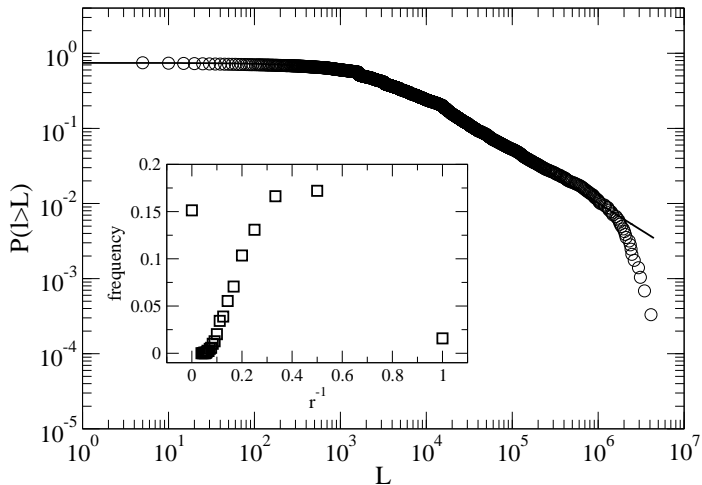}%
\caption{\label{load_fig} The probability that a substation has more than L transmission paths 
passing through it. The 
continuous curve has the generalized power law form (\ref{load_eq}). Inset: histogram of the length
of the shortest alternative path $r$ between the endpoints of an edge. In order to be able to include edges
with no alternative path, the abscissa is inverted. }
\end{figure}

As the role of the power grid is to transport power from generators to consumers, a possible measure 
for the importance of a node corresponding to a substation is its betweenness (or load)\cite{gkk01,n01}.
The betweenness of a node in a network is defined as the number of shortest paths that traverse 
it\cite{gkk01,n01}. Assuming that power is routed through the most direct path, the betweenness of a substation 
is a proxy for how much power it is transmitting, and for this reason we will use the alternative term 
load to denote it. Since it is the transmission substations' role to route power from generators 
to distribution substations, we focus our attention to them. We determine the shortest paths starting 
from all generation substations and ending on an all other reachable substations. For each transmission node we accumulate the number of paths that 
pass through it; being at the start or at the end of a path does not count. The highest possible load 
is $1633\times 12466\simeq 20$ million. We find that substations can have a load anywhere between 1 
and 4 million, and determine the cumulative load distribution, i.e. the probability that a node's load 
$l$ is larger than a given value $L$ (see Fig. \ref{load_fig}). The functional form of the 
cumulative load distribution is
\begin{equation}
P(l>L)\sim (2500+L)^{-0.7}
\label{load_eq}
\end{equation}
Fig. \ref{load_fig} illustrates that $40\%$ of the 
substations participate in tens or hundreds of paths only, but $1\%$ of them are part of a million 
or more paths. These high-load substations, although possibly not hubs regarding their degree, 
play an important role in power transmission. 

A fundamental requirement of the power grid is robustness, the ability to withstand and tolerate
errors (random failure) and targeted attacks\cite{ajb00,cebh01,cnsw00}. To ensure the reliability of 
power distribution, the transmission grid was conceived in such a way that there is more than one 
electrical path between any two points in the system \cite{power_book}. We wanted to verify whether the 
actual topology of the current power grid has this feature of global redundancy, or it has lost it during its 
growth and evolution. A possible measure of network redundancy is the so-called edge range, defined as 
the distance between the two endpoints of an edge if the edge connecting them were removed \cite{mnl02}.
 The inset of Fig. \ref{load_fig} shows the frequency of 
different edge ranges $r$ plotted as a function of $r^{-1}$. We find that parallel edges and short 
alternative paths are fairly frequent. However, around $15\%$ of the edges in the power grid have 
an infinite range. In addition to the $2179$ edges ending in distribution 
substations, close to $900$ edges connecting generators and/or transmission substations
are radial. These radial edges represent a clear vulnerability, as their loss disconnects 
their endpoints and creates isolated clusters in the power grid.

While the connectedness of the power grid allows for the transmission of power over large distances, 
it also implies that local disturbances propagate over the whole grid. The failure of a power line 
due to lightning strike or short-circuit leads to the overloading of parallel and nearby lines. 
Power lines are guarded by automatic devices that take them out of service when the voltage on them 
is too high. Generating substations are designed to switch off if their power cannot be transmitted; 
this protective measure has the unwanted effect of diminishing power for all consumers. Another 
possible consequence of power line failure is the incapacitation of transmission substations, 
possibly causing that the power from generators cannot reach distribution substations and ultimately 
consumers. 

In the unperturbed state each distribution substation can receive power from any of the
$N_g=1633$ generators. As substations lose function, the number of generators connected
to (and able to feed) a certain distribution substation $i$, $N_g^i$, decreases. We introduce
the concept of connectivity loss to quantify
the average decrease in the number of generators connected to a distributing substation,

\begin{equation}
CL=1-\left\langle\frac{N_g^i}{N_g}\right\rangle_i,
\end{equation}  
where the averaging is done over every distributing substation. In summary, the connectivity loss 
measures the decrease of the ability of distribution substations to receive power from the generators, and
in the following we will express it as a percentage.
 
\begin{figure}
\includegraphics[width=8cm,angle=-90]{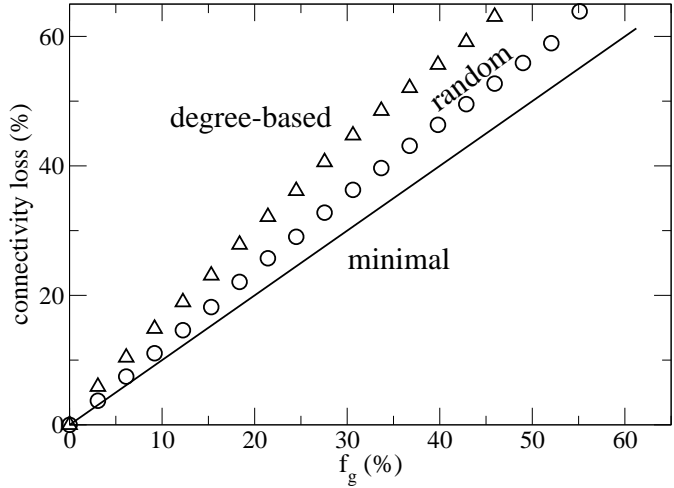}%
\caption{\label{source_fig} Connectivity loss in the power grid resulting from the failure of a 
fraction $f_g$ of generators. The straight line represents the minimum loss due to the 
node removal itself. Circles: random removal of generators; triangles: removal starting from the 
highest-degree generators. The curves are averages of ten runs, where either 
the list of generators or the list of generators with the same degree was randomly permuted.}
\end{figure}

 First we investigate the effect that the failure 
 of a power-generating substation has on consumers. Since initially the network contains a single
 connected component every consumer can reach all generators, and their connectivity is $100\%$. As 
 the number of generators decreases this value will decrease due to both loss of the generators themselves and 
 due to loss of routing capabilities at the generating substation level. 
 We remove nodes corresponding
 to generators either randomly, or in the decreasing order of their degrees, and monitor the connectivity loss
 as a function of the fraction of generators missing. The minimum possible loss is equal to the fraction 
 $f_g$ of inactive generators and is due to the loss in generation only (straight line on 
 Fig. \ref{source_fig}). We find that the 
 connectivity loss caused by removing power substations remains very close to this minimum value 
 (Fig. \ref{source_fig}), even though generating substations tend to be the 
 largest hubs in the system. The removal of generating substations does not alter the overall 
 connectivity of the grid thanks to a high level of redundancy at the power generating substation 
 level.
 
 \begin{figure}
\includegraphics[width=8cm,angle=-90]{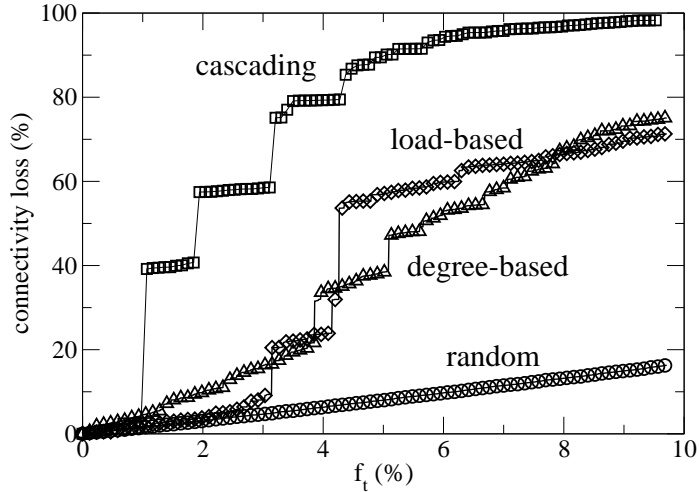}%
\caption{\label{trans_fig} Connectivity loss in the power grid due to the removal of nodes corresponding
to transmission substations. We remove a fraction $f_t$ of transmission nodes with four different
algorithms: randomly (circles), in the decreasing order of their degrees (triangles) or loads (diamonds),
and by recalculating the load every ten steps and removing the ten nodes with highest load (squares). The curves corresponding to 
random and degree-based node removal were averaged over ten runs. The load-based and cascading removal 
curves represent a single run.}
\end{figure}
 
The situation can be dramatically different when the nodes that we remove are transmission nodes. If 
the power grid were highly redundant the loss of a small number of transmission substations
should not cause power loss as power is rerouted through alternative paths. We find that even the removal
of a single transmission node causes a slight connectivity loss. We remove transmission nodes one by 
one,
first randomly, then in the decreasing order of their degree or load. 
For a random failure the connectivity loss is fairly low and stays proportional with the number of nodes 
lost. The connectivity loss is 
significantly higher, however, when targeting high degree or high load transmission hubs 
(Fig. \ref{trans_fig}). The grid can withstand only a few failures of this nature before considerable parts of the network
become disconnected leading to substantial connectivity loss at consumer level. For example,
failure of only $4\%$ of the nodes with high load may cause up to $60\%$ loss of connectivity. We also study an algorithm where we periodically recalculate the
load of all transmission nodes during node removal, and select the nodes with highest load to
be deleted next. This is a possible illustration of a propagating (cascading) power failure, 
where it is more likely that substations that have the highest load in the perturbed configuration
will fail next. Fig. \ref{trans_fig} illustrates that this cascading failure has the most damaging effect, 
as the loss of only $2\%$ of the high-load transmission substations leads to a connectivity loss of
almost $60\%$, and all distribution substations become virtually powerless at $f_t \simeq 8\%$. In conclusion,
the transmission hubs ensuring the connectivity of the power grid are also its largest liability in 
case of power breakdowns.

This vulnerability of the electric power grid is inherent to its organization and therefore cannot 
be easily addressed without significant investment. Possible solutions include increasing the 
redundancy and capacity of the existent structure or decreasing the reliance on transmission by 
incorporating more generation at the distribution substation level. Such distributed generation 
by small local plants can supplement power from the grid under normal operation conditions and 
can greatly mitigate the effects of blackouts on the population. Targeted use of generation 
located near the point of use might prove to be the only viable economical alternative.

\begin{acknowledgements}
 The authors wish to thank Donna Heimiller and Steven Englebretson for 
their help in obtaining the POWERmap network data. This research was partially supported 
by the Midwest Research Institute (contract number AAX-3--33641-01).
\end{acknowledgements}


\begin{thebibliography}{10}
\bibitem{roadmap}
Electricity Technology Roadmap, 1999 Summary and Synthesis, by the Electric Power Research 
Institute, \url{http://www.epri.com/corporate/discover_epri/roadmap/}
\bibitem{rely}
North American Electricity Reliability Council reliability assessment report, 1998,
\url{http://www.nerc.com/~filez/rasreports.html}
\bibitem{flow}
Dromey Design electrical distribution analysis software, 
\url{http://www.dromeydesign.com/dess/lfa.htm}
\bibitem{ab02}
R. Albert and A.-L.  Barab\'asi , {\it Reviews of Modern Physics} {\bf 74}, 44-94 (2002); A.-L. Barab\'asi
{\it Linked: The New Science of Networks} (Perseus Publishing, Cambridge, 2002); D. J. Watts 
{\it Six Degrees: The Science of a Connected Age} (W. W. Norton $\&$ Co., New York, 2003); S. N. Dorogovtsev 
and J. F. F. Mendes, {\it Evolution of Networks: From Biological Nets to the Internet and WWW} (Oxford
University Press, Oxford, 2003); M. E. J. Newman, {\it SIAM Review} {\bf 45}, 167 (2003).
\bibitem{platts}
Platts Global Energy, \url{http://www.platts.com/electricpower/index.shtml}.
\bibitem{miss}
This method may miss distribution substations with more than one incoming transmission line. 
Unfortunately no information regarding the directionality of the transmission lines was available
from our data sources.
\bibitem{ajb00}
R. Albert, H. Jeong and A.-L. Barab\'asi, {\it Nature} {\bf 406}, 378 (2000).
\bibitem{cebh01}
 R. Cohen,  K. Erez, D. ben-Avraham  and S. Havlin, {\it Phys. Rev. Lett.} {\bf 85}, 4626
(2000); R. Cohen,  K. Erez, D. ben-Avraham  and S. Havlin, {\it Phys. Rev. Lett.} {\bf 86}, 3682 (2001).
\bibitem{cnsw00}
D. S. Callaway, M. E. J. Newman, S. H.  Strogatz and D. J. Watts,  {\it Phys. Rev. Lett.}
 {\bf 85}, 5468 (2000).
\bibitem{asbs00}
L. A. N. Amaral, A.  Scala, M. Barth\'el\'emy  and H. E. Stanley, {\it  Proc. Natl. Acad. Sci. 
USA} {\bf 97}, 11149 (2000).
\bibitem{gkk01}
K.-I. Goh, B. Kahng and  D.  Kim, {\it Phys. Rev. Lett.} {\bf 87}, 278701 (2001).
\bibitem{n01}
M. E. J. Newman,  {\it Phys. Rev. E} {\bf 64}, 016132 (2001).
\bibitem{power_book}
H. Saadat, {\it Power System Analysis} (McGraw-Hill, Boston, 1999).
\bibitem{mnl02}
A. E. Motter, T. Nishikawa and Y.-C. Lai, {\it Phys. Rev. E} {\bf 66}, 0651103(R), (2002).
\end{thebibliography}
\end{document}